\documentclass[%
 preprint,
superscriptaddress,
 amsmath,amssymb,
 aps,
 longbibliography
]{revtex4-2}

\usepackage{graphicx}
\usepackage{dcolumn}
\usepackage{bm}
\usepackage{amsfonts}
\usepackage{amssymb}
\usepackage{amsthm,amsmath}
\usepackage{mathrsfs}
\usepackage{indentfirst}
\usepackage{multirow}
\usepackage{subfigure}
\usepackage{upgreek}
\usepackage{hyperref}
\hypersetup{hypertex=true,
            colorlinks=true,
            linkcolor=blue,
            anchorcolor=blue,
            citecolor=blue}
\UseRawInputEncoding
\begin{document}

\title{Contact angle measurement on curved wetting surface in multiphase lattice Boltzmann method}

\author{Yangsha Liu}
\author{Yichen Yao}
\author{Quanying Li}
\author{Binghai Wen}
\email{Corresponding author. Email:oceanwen@gxnu.edu.cn}
\affiliation{%
 Guangxi Key Lab of Multi-Source Information Mining \& Security, Guangxi Normal University, Guilin 541004, China
}%
\affiliation{%
 School of Computer Science and Engineering, Guangxi Normal University, Guilin 541004, China
}%

\date{\today}
\begin{abstract}
Contact angle is an essential physical quantity that characterizes the wettability of a substrate. Although it is widely used in the studies of surface wetting, capillary phenomena and moving contact lines, measuring contact angles in experiments and simulations is still complicated and time-consuming. In this paper, we present an efficient scheme for the real-time and on-the-spot measurement of contact angles on curved wetting surfaces in lattice Boltzmann simulations. The measuring results are in excellent agreement with the theoretical predictions by the spherical cap method without considering the gravity effect. A series of the simulations with various drop sizes and surface curvatures confirm that the present scheme is grid-independent. Then, it is verified in gravitational environments by simulating the deformations of sessile and pendent droplets on the curved wetting surface. The numerical results are highly consistent with experimental observations and support the theoretical analysis that the microscopic contact angle is independent of gravity. Furthermore, the scheme is applied to capture the dynamic contact angle hysteresis on homogeneous or chemically heterogeneous curved surfaces. Importantly, the accurate contact angle measurement enables the mechanical analysis at moving contact lines. The present measurement is simple and efficient, and can be extended to implement in various multiphase lattice Boltzmann models. 
\end{abstract}

\maketitle

\section{\label{sec1}INTRODUCTION}
Contact angle is an important characteristic quantity used to express surface wettability and has a wide range of applications in nature and industrial production, such as wetting, microfluidics, capillary phenomena, coating technology and moving contact line 
\cite{Andreotti2020,Sui2014,Snoeijer2013}. Experimenters have researched and invented various schemes in order to measure contact angles. One of the earliest and widely used methods is the technique of measuring the contact angle of sessile droplets using a telescopic goniometer, which is based on the principle of measuring the line tangency of the three-phase contact points of a droplet profile on a smooth surface \cite{Bigelow1946}. The angle measured in this way is usually very close to advancing contact angles. By applying explicit vibration \cite{DellaVolpe2020}, equilibrium contact angles can be obtained. McDougall \emph{et al.} modified the sessile drop method and obtained the advancing and receding contact angles by tilting the solid surface until the droplet just started to move \cite{Macdougall1942}. Subsequently, Extrand and Kumagai used this method to study the contact of liquids on various polymer surfaces \cite{Extrand2003}. Axisymmetric drop shape analysis-profile is a technique used to measure liquid-fluid interfacial tension and contact angles, and has high precision \cite{Cheng1990}. Kwok \emph{et al.} control droplet injection rate or extraction rate and then use the technique to measure low-velocity dynamic contact angle \cite{Kwok1998}. Langmuir and Schaeffer used the specular reflection of the droplet surface to measure the contact angle \cite{Langmuir1937}. Later Fort and Patterson improved the method and used it for static drops and meniscus on flat plates or inner tubes \cite{Fort1963}. Different from the methods where a sessile droplet is formed above a solid sample, the captive bubble  method provides a direct measurement of the contact angle of bubble formation in a liquid by forming a bubble below the solid sample and then immersing it in the test liquid \cite{Zhang1990}. The method of tilting the plate is to immerse one end of the solid plate in the liquid, and rotate the other end toward the liquid surface until it is immersed in the liquid, forming the meniscus on both sides of the plate \cite{Bezuglyi2001}. The plate is slowly tilted until the meniscus on one side becomes horizontal. The angle between the plate and the horizontal plane is the contact angle. In addition, researchers are not only able to calculate the contact angle by direct measurement, but also by indirect methods. In the Washburn capillary rise method, for example, the contact angle is derived from the rate at which the liquid rises through the powder-filled bed by capillary action \cite{Washburn1921}. The Wilhelmy balance method is also one of the common methods for measuring contact angles \cite{Tretinnikov1994}, which is an indirect force method that reduces the measurement of angle to a measurement of weight and length. The result of this method is highly accurate and not subjective, and it is also suitable for studying the advancing or receding contact angle and contact angle hysteresis.

In numerical simulations involving surface wettability, the scheme that used a goniometer to measure the contact angle from images generated by the simulated data is subjective and rough. More precisely, image analysis can be employed to obtain the contact angle from images \cite{Bommer2014}. Sakugawa \emph{et al.} obtained contact angles by using image processing and polynomial fitting \cite{Sakugawa2020}. In order to improve the accuracy of contact angle, Scanziani \emph{et al.} and Klise \textit{et  al}. both used X-ray microtomography images to calculate the contact angle \cite{Klise2016,Scanziani2017}.  Measuring contact angles in low-resolution images is cumbersome, especially as the need to derive fluid images prior to measurement is tedious and time-consuming, and may introduce subjective bias, so is not optimal for in-situ measurements. Without the influence of gravity, surface tension makes the droplet appear as a spherical cap shape on a plain surface. The contact angle can be calculated by measuring the height and bottom width of the droplet \cite{Huang2007,Chen2014} , and this theoretical method is known as the spherical cap method. When the droplet is reduced to the nanometer scale, since there is no stable interface between gas and liquid, the descending contour needs to be fitted by the least square method \cite{Wang2009,Wang2015}. The spherical cap method is simple and achievable, but it cannot be used in gravity or nonequilibrium environments. Subsequently, researchers have gradually expanded the study of contact angles from droplets to fibres and porous media, and have proposed several methods to measure their contact angles from an energy perspective. Amrei \emph{et al.} studied the variation of rough fibre contact angle with fibre roughness by means of an energy minimization method \cite{Amrei2017}. Blunt \emph{et al.} determined the contact angle of three-phase flow in porous media by using energy balance \cite{Blunt2021}. Jasper proposed a general variational method for predicting contact angles considering the Laplace pressure case \cite{Jasper2019}. However, these are domain-specific methods and are not generic.

There are also many studies in numerical simulation on the modification of boundary conditions to improve the accuracy of contact angle measurements. For the simulation of diffusion interfaces, prescribed contact angles can be obtained by using geometric formulas for the wetting conditions \cite{Ding2007}. To improve the accuracy and stability of the contact angle boundary conditions, Lee \emph{et al.} use characteristic interpolation to obtain contact angles \cite{Lee2011}. Dong further extended the boundary conditions of the contact angle after considering the relaxation of the dynamic contact angle to simulate dynamic wall-confined gas/liquid flows with large density ratios \cite{Dong2012}. Leclair \emph{et al.} used Dirichlet boundary conditions to study incompatible two-phase pore-scale suction and discharge forces in porous media using the desired contact angle imposed at the boundary \cite{Leclaire2016,Leclaire2017}. These methods aim to modify the wetting boundary condition to impose an accurate contact angle, rather than improve the algorithm of the contact angle measurement. These imposing procedures of contact-angle boundary conditions are computationally complex and nonlocal. Especially, they involve the intervention to the evolution of flow field. Essentially, a contact angle is a geometrical concept. The Young’s equation can only theoretically explain some special cases, such as a sessile drop on a flat substrate at zero-gravity mechanical equilibrium. In dynamic or nonequilibrium environments, the contact angle should be measured through a geometrical method. Recently, Wen \emph{et al.} proposed a geometry-based contact angle measurement on a plain substrate. The simulation results showed that the method was accurate and efficient \cite{Wen2018}. Nevertheless, experiments, natural phenomena and industrial applications often involve complex boundary shapes and even soft substrates, which appeal to an effective and real-time scheme to measure the contact angle on curved surfaces.

In this paper, we design a in situ method for contact angle measurement on curved wetting surface. In Section \ref{sec2}, we introduce the lattice Boltzmann method and the chemical-potential multiphase model. Section \ref{sec3} describes in detail the measuring method contact angle on curved wetting surface and the chemical-potential boundary condition. In Section \ref{sec4}, we verify the measurement accuracy and the grid-independence. A series of simulations of sessile and pendent droplets under the effect of gravity indicate that the contact angle is microscopic and independent of gravity. Subsequently, the dynamic hysteresis phenomena of contact angle of droplets on homogeneous surfaces as well as chemically patterned surfaces are computed and analyzed. Based on the accurate contact angle, we can perform the mechanical analyses at the contact line regions of droplets on chemically patterned surfaces. Finally, Section \ref{sec5} briefly summarizes the work.

\section{\label{sec2}Multiphase lattice Boltzmann method}

\subsection{\label{sec2.1}Lattice Boltzmann method}
Lattice Boltzmann (LB) method has developed into a very effective numerical method for simulating complex fluid flow \cite{Chen1998,Aidun2010,He2019,Chai2012,Zhang2017,Gan2012,Liu2013}. LBM is derived from the concept of cellular automata and kinetic theory, and its inherent mesoscopic properties make it excellent in modeling fluid systems involving interface dynamics \cite{Ladd2001,Wen2014,Wen2015a} and phase transitions \cite{Chen2014,Li2016}. The lattice Boltzmann equation (LBE) is fully discretized in space, time, and velocity. The multiple-relaxation-time (MRT) version of LBE improves the numerical stability and computational accuracy, and it can be expressed as \cite{Lallemand2000}
\begin{equation}
    {f_i}({\bm{x}} + {{\bm{e}}_i}\delta t,t + \delta t) - {f_i}({\bm{x}},t) =  - {{\rm{M}}^{ - 1}} \cdot {\rm{S}} \cdot [{\rm{m}} - {{\rm{m}}^{({\rm{eq}})}}] + {F_i}\label{eq(1)}
\end{equation}
where  $\bf{M}$ is a transformation matrix that linearly transforms the distribution functions to the velocity moments; $\bf{m}$  and $\bf{m}^{({\rm{eq}})}$  represent the velocity  moments of the distribution functions and their equilibria, ${\bf{m}} = {\bf{M}} \cdot {\bf{f}}$ ; and ${\bf{f}} = {{\bf{M}}^{ - 1}} \cdot {\bf{m}}$, where ${\bf{f}} = \left( {{f_0},{f_1}, \ldots ,{f_8}} \right)$  for the D2Q9 model. ${f_i}({\bm{x}},t)$  is the particle distribution function at time  $t$ and lattice site ${\bm{x}}$ , moving along the direction defined by the discrete velocity vector ${{\bm{e}}_i}$  with $i = 0, \ldots ,N$ . $f_i^{(eq)}$  is the equilibrium distribution function
\begin{equation}
    f_{i}^{(eq)}(\bm{x},t)=\rho {{\omega }_{i}}\left[ 1+3\left( {{\bm{e}}_{i}}\cdot \bm{u} \right)+\frac{9}{2}{{\left( {{\bm{e}}_{i}}\cdot \bm{u} \right)}^{2}}-\frac{3}{2}{{\bm{u}}^{2}} \right]\label{eq(2)}
\end{equation}
where  ${\omega _i}$ is the weighting coefficient and ${\bm{u}}$  is the fluid velocity.

The lattice Boltzmann Eq. \eqref{eq(1)} is decomposed into two basic steps of collision and advection, revealing the phenomenon of fluid movement at the meso level.
\begin{gather}
  {\rm{collision: }}{f_i}({\bm{x}},t) = {f_i}({\bm{x}},t) - \frac{1}{\tau }[ {{f_i}({\bm{x}},t) - f_i^{(eq)}({\bm{x}},t)}\label{eq(3)}\\
  {\rm{advection: }}{f_i}\left( {{\bm{x}} + {{\bm{e}}_i},t + 1} \right) = {\tilde f_i}({\bm{x}},t)\label{eq(4)}
\end{gather}

In the MRT model, its biggest feature is that multiple relaxation times are used in the collision process, and different moments can use different relaxation times. ${\bf{S}}$ is a diagonal matrix of non-negative relaxation times: ${\bf{S}} = {\mathop{\rm diag}\nolimits} \left( {0,{s_e},{s_\varepsilon },0,{s_q},0,{s_q},{s_v},{s_v}} \right)$ . In this paper, the relaxation times are given by ${s_e} = 1.64$, ${s_\varepsilon } = 1.54$,  ${s_q} = 1.9$, ${s_v} = 1/\tau $ for the simulations with the MRT LBE.
\subsection{\label{sec2.2}Chemical-Potential multiphase model}
The chemical potential is the partial differential of the Gibbs free energy to the composition \cite{Jamet2002}. For a nonideal fluid system, following the classical capillarity theory of van der Waals, the free energy functional within a gradient-squared approximation is written as \cite{rowlinsonmolecular,Swift1995}
\begin{equation}
    \Psi =\int{\left[ \psi (\rho )+\frac{\kappa }{2}|\nabla \rho {{|}^{2}} \right]}d{x}\label{eq(5)}
\end{equation}
where the first term represents the bulk free-energy density and the second term describes the contribution from density gradients in an inhomogeneous system, and $\kappa $ is the surface tension coefficient. The general chemical potential can be derived from the free energy density functional \cite{Jamet2002,Zheng2006},
\begin{equation}
    \mu ={\psi }'(\rho )-\kappa {{\nabla }^{2}}\rho \label{eq(6)}
\end{equation}
Gradients in the chemical potential act as a thermodynamic force on the fluid. With respect to the ideal gas pressure $c_s^2\rho $, the nonideal force can be evaluated by a chemical potential 
\begin{equation}
    \bm{F}=-\rho \nabla \mu +c_{s}^{2}\nabla \rho \label{eq(7)}
\end{equation}
The general equation of state can also be defined by the free energy density,
\begin{equation}   
    {{p}_{0}}=\rho {\psi }'(\rho )-\psi (\rho )\label{eq(8)}    
\end{equation}
Solving the linear ordinary differential Eq. \eqref{eq(8)} gives the general solution of the free-energy density
\begin{equation}
    \psi =\rho (\int{\frac{{{p}_{0}}}{{{\rho }^{2}}}d\rho }+C) \label{eq(9)}
\end{equation}
where  $C$ is a constant. When the general expression of equation of state (EOS) is selected, substituting Eq. \eqref{eq(9)} into Eq. \eqref{eq(8)} will solve the relevant chemical potential, and the constant is eliminated. For example, the famous Peng-Robinson (PR) EOS and its chemical potential are,
\begin{equation}
    {{p}_0}=\frac{\rho RT}{1-b\rho }-\frac{a\alpha (T){{\rho }^{2}}}{1+2b\rho -{{b}^{2}}{{\rho }^{2}}} \label{eq(10)}
\end{equation}
and 
\begin{equation}
    \mu _{{}}^{\text{PR}}=RT\ln \frac{\rho }{1-b\rho }-\frac{a\alpha (T)}{2\sqrt{2}b}\ln \frac{\sqrt{2}-1+b\rho }{\sqrt{2}+1-b\rho }+\frac{RT}{1-b\rho }-\frac{a\alpha (T)\rho }{1+2b\rho -{{b}^{2}}{{\rho }^{2}}}-\kappa {{\nabla }^{2}}\rho  \label{eq(11)}
\end{equation}
where $R$ is the gas constant, $a$ is the attraction parameter, $b$ is the volume correction parameter, and the temperature function is \begin{small}$\alpha (T) = {\!\left[ {1\! + \!\left( {0.37464\! +\! 1.54226\omega\!  -\! 0.26992{\omega ^2}} \right)\left( {\!1 \!- \!\sqrt {T/{T_c}} } \right)} \right]^2}$\end{small}. In our simulations, the parameters are given by $a = 2/49$,  $b = 2/21$, and $R = 1$. The acentric factor $\omega$ is 0.344 for water. To make the numerical results closer to the actual physical properties, we define the reduced variables ${T_r} = T/{T_c}$  and ${\rho _r} = \rho /{\rho _c}$  , in which  ${T_c}$ is the critical temperature and  ${\rho _c}$ is the critical density.

A proportional coefficient  $k$ is introduced to decouple the dimension unit of the length between the momentum space and the mesh space, namely $\delta \hat x = k\delta x$ . Here the quantities in the mesh space are marked by a superscript. Following dimensional analysis, the chemical potential in the mesh space can be evaluated by \cite{Wen2017,Wen2020} 
\begin{equation}
        \hat{\mu }={{k}^{2}}{\psi }'(\rho )-\hat{\kappa }{{\hat{\nabla }}^{2}}\rho\label{eq(12)}
\end{equation}                   	                             
We further apply the central difference method with fourth order accuracy to calculate the gradients. These approaches greatly improve the stability of the chemical-potential multiphase model, and the transformation has no loss of accuracy holding the mathematical equivalence. 

In addition, we chose the exact difference method proposed by Kupershtokh \emph{et al.} to incorporate the nonideal force  ${\bm{F}}$ into LBE \cite{Kupershtokh2009}:
\begin{equation}
    {{F}_{i}}=f_{i}^{(eq)}(\rho ,\bm{u}+\delta \bm{u})-f_{i}^{(eq)}(\rho ,\bm{u})\label{eq(13)}
\end{equation} 

where $\delta {\bm{u}} = \delta t{\bm{F}}/\rho $ . The body force term  ${F_i}$ is simply equal to the difference of the equilibrium distribution functions before and after the nonideal force acting on the fluid during
a time step. Correspondingly, the macroscopic fluid velocity is redefined as the average momentum before and after the collision: ${\bm{v}} = {\bm{u}} + \delta t{\bm{F}}/(2\rho )$ .

\section{\label{sec3}Mesoscopic measurement of contact angle on curved surface}
\subsection{\label{sec:3.1}Chemical-potential boundary condition}
\begin{figure}[]
	\centering
		\includegraphics[scale=.75]{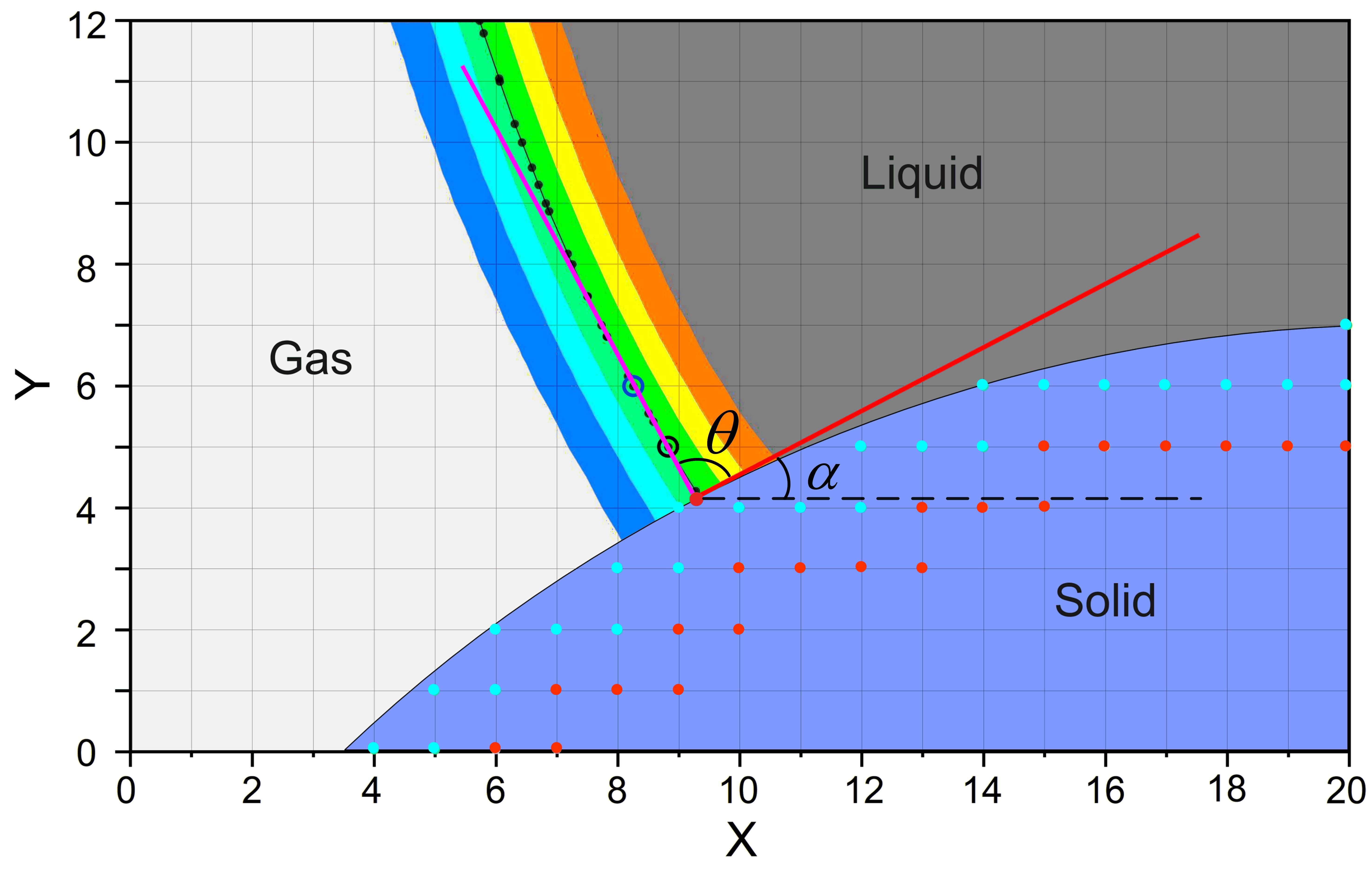}
	\caption{A schematic diagram of three-phase contact region of a drop on a curved substrate. The black curve (in rainbow region) represents the liquid-gas interface of the drop, and the black points mark the intersections of the interface and liquid-gas links. The two points marked by black and blue circles are approximate 1 and 2 lattice units away from the substrate, and the pink line passing the two points intersects the substrate at the red point, which is used as the three-phase contact point. The contact angle $\theta $ is defined by the pink line and the tangent line of the curved substrate at the three-phase contact point, and $\alpha $ is the inclination angle of the tangent line relative to the horizontal line. }
	\label{FIG:1}
\end{figure}

The chemical potential plays a fundamental role in driving a phase transition or indicating the wettability of a solid surface. In the previous work, the chemical-potential boundary condition is implemented on a plane surface \cite{Wen2018,Wen2020,He2020}. Here, we improve the scheme to a curved wetting surface. Fig. \ref{FIG:1} presents the three-phase contact region of a drop on a curved wetting substrate. Since the central difference method with fourth order accuracy involves the neighboring nodes whose distances are less than or equal to two lattice units, the chemical-potential boundary condition must treat two layers of solid nodes at least. A specific chemical potential is assigned to these solid nodes in order to regulate the wettability of the substrate. It influences on the gradient calculation of chemical potential on the fluid nodes adjacent to the substrate, and this chemical-potential gradient reflects the interaction between the fluid and the wetting substrate. On the other hand, the boundary condition needs to estimate the densities of the solid nodes in the two layers in order to calculate the density gradient. Fig. \ref{FIG:1} marks the first and second layers of solid lattice nodes by cyan and orange, respectively. The densities on the two layers of solid nodes can be calculated based on the nearest neighbor nodes,

\begin{equation}
    \rho \left( {{\bm{x}}_{s}} \right)=\frac{\sum\limits_{i}{{{\omega }_{i}}}\rho \left( {{\bm{x}}_{s}}+{{\bm{e}}_{i}}{{\delta }_{t}} \right){{s}_{w}}}{\sum\limits_{i}{{{\omega }_{i}}}{{s}_{w}}}\label{eq(14)}
\end{equation} 
where  $\bm{x_s} + {{\bm{e}}_i}{\delta _t}$ indicates the adjoining nodes, and ${s_w}$ is a switching function. For the first layer of nodes (in cyan), ${s_w} = 1$  when $\bm{x_s} + {{\bm{e}}_i}{\delta _t}$  is a fluid node; for the second layer of nodes (in orange), ${s_w} = 1$  when $\bm{x_s} + {{\bm{e}}_i}{\delta _t}$ is in the first layer; otherwise,  ${s_w} = 0$ . The miss distribution functions on the boundary, which stream, in concept, from a solid node to a fluid node, are calculated by the multiphase curved boundary condition with mass conversation \cite{RN5164}. 

\subsection{\label{sec3.2}Contact angle measurement on curved wetting surface}
In natural phenomena and scientific researches, the wetting substrates often have complex boundary shapes. The contact angle measurement on these curved wetting surfaces is very useful to depict the phase transition and contact line moving. Especially, in recent years, the advances in elastic capillarity and soft matter have brought soft wetting to the attention of scientists \cite{Andreotti2020}. Where localized deformations of soft materials occur, the contact angle measurement are highly desirable for the local mechanical analyses.  As shown in Fig. \ref{FIG:1}, the liquid-gas interface is defined by the contour line where the density is equal to the mean density of the gas and liquid phase, which is very close to the theoretical interface defined by the equimolar division and has a much simpler calculation \cite{Ladd2001}. This surface distinguishes the liquid and gas nodes in the transition region of the drop. The intersections between the drop surface and the liquid-gas links can then be obtained by the linear interpolation,
\begin{equation}
   \bm{x} = \bm{x_g} + \frac{{{\rho _m} - {\rho _{({x_g})}}}}{{{\rho _{({x_l})}} - {\rho _{({x_g})}}}}\bm{e}\label{eq(15)}
\end{equation} 
where $\bm{x_l}$  and $\bm{x_g}$ represent the liquid and gas nodes of a liquid-gas link respectively, and  ${\rho _m}$ is the mean density of the liquid and gas. The two intersections, whose distances are approximate 1 and 2 lattice units away from the curved substrate, are marked by the black and blue circles, respectively. The pink line passing the two points intersects the drop surface, and the intersection is defined as the three-phase contact point. The contact angle $\theta $ is determined by the red line and the tangent line of the curved substrate at the three-phase contact point.
\begin{figure}[]
	\centering
		\includegraphics[]{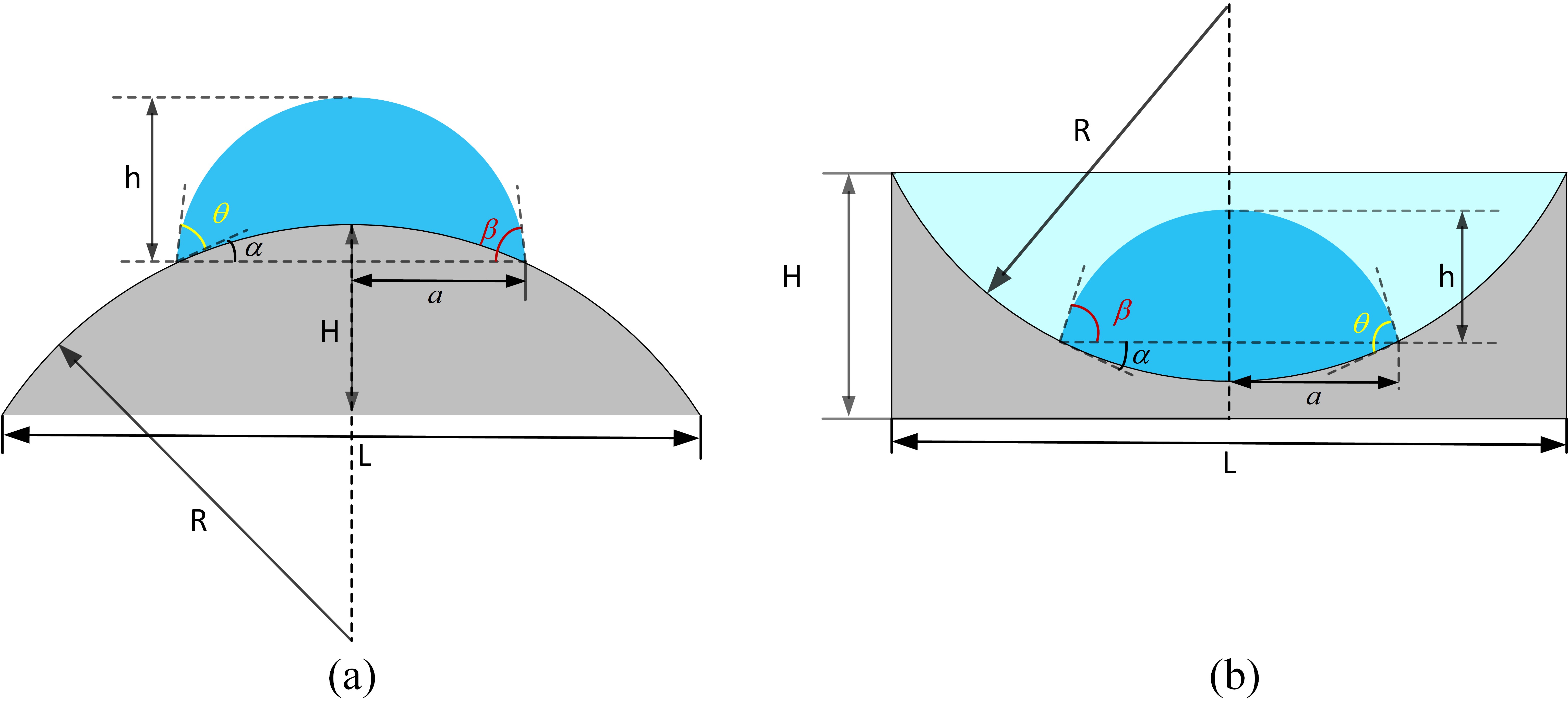}
	\caption{Schematic diagrams of a droplet on (a) convex or (b) concave wetting surfaces.}
	\label{FIG:2}
\end{figure}

Fig. \ref{FIG:2} presents a droplet on convex or concave wetting substrates. The curved substrate is parameterized by the width $L$, the height $H$ and the curvature radius $R$. The letters $a$ and $h$ indicate the contact radius and height of the drop, respectively. The inclined angle of the drop interface with respect to the horizontal line is $\beta $, which is equal to $\theta  + \alpha $ for a convex surface and $\theta  - \alpha $ for a concave surface.

\section{\label{sec4}Simulation and discussion}
In this section, we perform a series of numerical simulations to demonstrate the effectiveness of the contact angle measurement on curved wetting surfaces. At first, in an environment without gravity, the measuring accuracy on the substrates from hydrophilic to superhydrophobic is verified by comparing with the benchmarks computed by the spherical cap method. The mesh independence is further confirmed, because the contact angles measured from drops with different sizes remain the same. Then, with the gravity effect, we verify the theoretical prediction that the microscopic contact angle is independent with gravity by simulating the deformations of two sets of sessile and pendent drops on the curved substrates. The dynamic contact angle hysteresis is captured on-the-spot in the simulations of a drop rolling on a curved and chemical-heterogeneous surface. These support that what the present scheme measures is the microscopic contact angle and the measurement is real-time. Because the microscopic contact angle reflects the mechanical equilibrium at the three-phase contact region, we finally perform the in-situ mechanical analysis during the drop movement on a curved wetting substrate. 

The droplet radius is ${r_0} = 40$ lattice units. Under gravity-free conditions, the computational domain  is a rectangle with the length 700 and width 400 lattice units, and the relationship between contact angle and chemical potential is investigated at two temperatures. The same flow field is then used to verify the grid-independence, and the deformation of the sessile droplets and the pendent droplets under the influence of gravity is further simulated. The temperature is $Tr = 0.6$ . The droplet density is $1{\rm{ }}g/c{m^3}$  and the gravitational acceleration is  $\left| {\left. G \right|} \right. = 980{\rm{ }}cm/{s^2}$. The droplet on the lattice unit is mapped onto the macroscopic droplet by the dimensional transformation. As the macroscopic droplet size increases, the gravitational effect becomes more and more obvious. After 100,000 time steps of free evolution, gravity gradually acts on the fluid (both gas and liquid) and finally reaches the equilibrium state. To better capture the dynamic hysteresis and real-time mechanical analysis of droplets on surfaces, the calculation field $Dx$ is extended to 3000 lattice units and the droplet radius is ${r_0} = 100$ lattice units. In this paper, droplets on a curved solid surface with a specific chemical potential are simulated using the PR EOS. The density of the flow field is initialized as follows \cite{Huang2011}:
\begin{equation}
   \rho (x,y) = \frac{{{\rho _g} + {\rho _l}}}{2} + \frac{{{\rho _g} - {\rho _l}}}{2}\tanh [\frac{{2(r - {r_0})}}{W}]\label{eq(16)}
  \end{equation}
where ${\rho _g}$  and ${\rho _l}$  are the gas and liquid coexistence densities obtained using Maxwell’s equal area method of construction, the initial interface width is $W = 10$ ,  ${r_0}$ is the initial radius of the droplet, and $r = \sqrt {{{(x - {x_0})}^2} + {{(y - {y_0})}^2}}$.

\subsection{\label{sec4.1}Accuracy of contact angle measurement}
\begin{figure}[]
	\centering
		\includegraphics[]{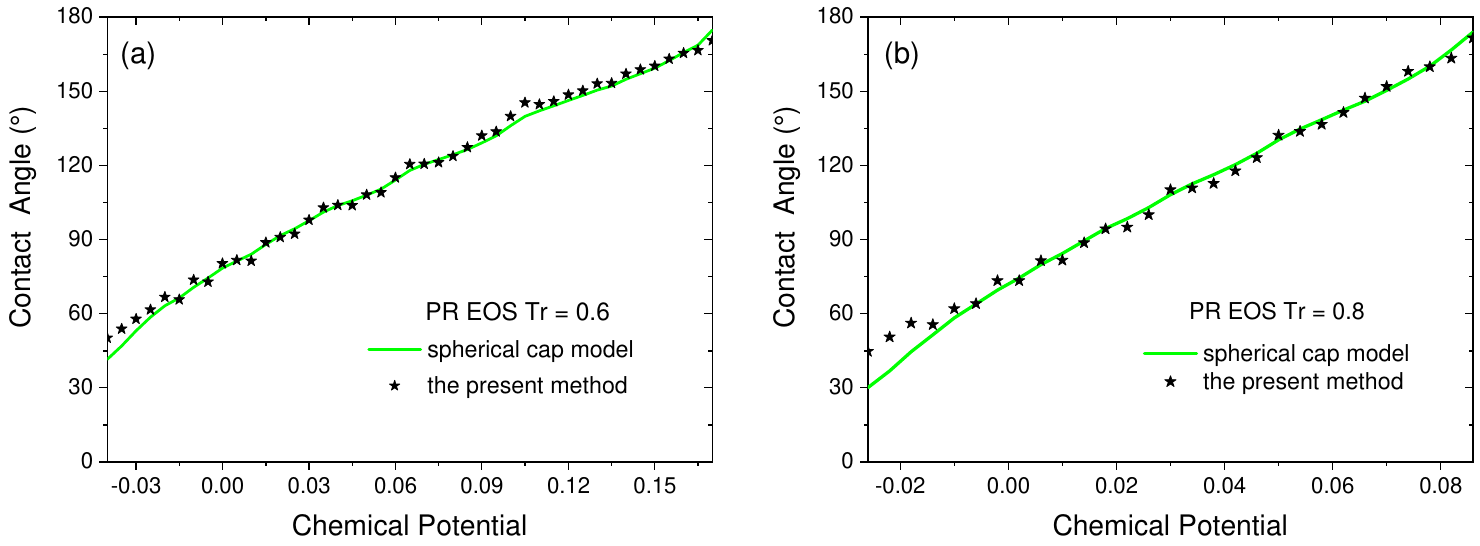}
	\caption{The contact angle measurements on a curved wetting surface by the present scheme and the spherical cap method at the temperatures (a) Tr = 0.6, and (b) Tr = 0.8.}
	\label{FIG:3}
\end{figure}

The droplet will have a perfect spherical cap under gravity-free conditions. If the length and height of the bottom of the droplet are $L$ and $H$, then the radius of the droplet can be calculated ${R_{\rm{0}}} = (4{H^2} + {L^2})/8H$ and then the horizontal angle  $\tan \beta  = L/2({R_0} - H)$, and then the contact angle can be calculated by the spherical cap method. The spherical cap method is generally used as a benchmark to validate the proposed measurement method for calculating contact angles on a mesoscopic scale, as the base length and height of the droplet can be readily calculated. Fig. \ref{FIG:3} shows the contact angles measured by the present method and the spherical cap method at two temperatures from hydrophilic to superhydrophobic surfaces. It can be seen from the Fig. \ref{FIG:3} that the results of the present method (black star) agree with those of the spherical cap method (green line). The overall trend for both the present method and the spherical cap method is linear, with the contact angle increasing as the chemical potential increases, but the spherical cap method is intuitively seen to bend in the superhydrophilic and superhydrophobic cases. We therefore fitted the linear type by least squares for the accuracy of this scheme.

To investigate the accuracy of the present scheme at temperature and for different wettability, the contact angles are measured on the hydrophilic and hydrophobic surfaces, and are drawn in Fig. \ref{FIG:3} as a function of chemical potential at the temperatures of 0.6 and 0.8. The contact angles calculated by the present scheme are in good agreement with the results by the spherical cap method. Nevertheless, the contact angles calculated by the spherical cap method show clearly bent when it is less than 60° or larger than 160°, whereas those from the present scheme keep a nice linear relationship with chemical potential of the surface. We further analyze the results quantitatively by the relative ${{\rm{L}}_{\rm{2}}}$-norm error, which is defined as  $E = \frac{{{{\{ \int {{{[f(t) - F(t)]}^2}dt} \} }^{1/2}}}}{{{{\{ \int {[F{{(t)}^2}dt]} \} }^{1/2}}}}$  , where  $f(t)$ is the result of the present scheme and  $F(t)$ is the linear fit by the least squares method. The measuring results at the temperature 0.6 and 0.8 have fairly small errors of 0.025 and 0.018, respectively. These manifest that the present scheme is accurate and stable. It is noteworthy that the linear relationship between the contact angle and the surface chemical potential is very useful in multiphase simulations, because the surface wettability can be readily adjusted according to actual requirements.

\subsection{\label{sec4.2}Verifications of grid independence}
\begin{figure}[]
	\centering
		\includegraphics[]{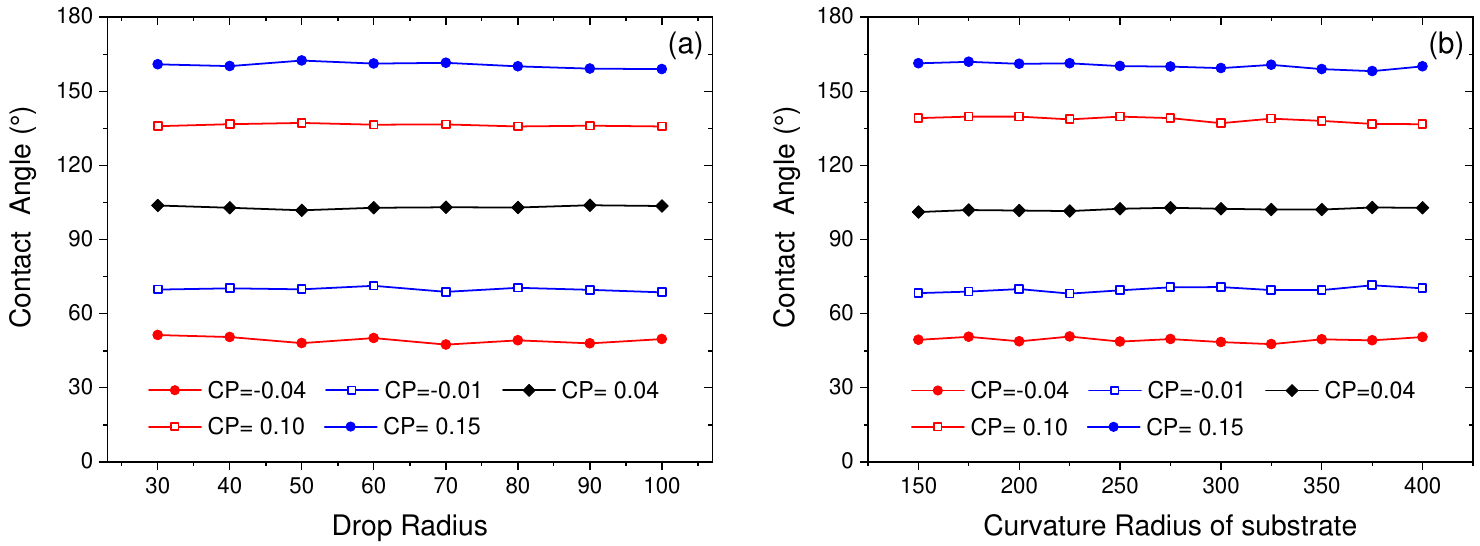}
	\caption{Verifications of grid independence on various curved wetting substrates. (a) The contact angles measured by the present scheme are independent of the drop sizes. (b) The contact angles measured by the present scheme are independent of the curvature radii of the substrates.}
	\label{FIG:4}
\end{figure}

Verification of the grid-independence is crucial in the results of numerical simulations. We first measure the contact angles of a series of droplets from   ${r_0} = 30$ to ${r_0} = 100$  lattice units on five different wettability substrates and find that the contact angles are the same for different droplet sizes. On the other hand, we measure the contact angles of droplets on substrates with the curvature radii from 150 to 400 lattice units. As can be seen in Fig. \ref{FIG:4}, the contact angles measured for different droplet sizes are the same and the contact angles for the same wettability on substrates with different radii of curvature are highly consistent. Therefore, neither the droplet size nor the radius of curvature of the substrate affects the contact angle measurement. Further quantitative analysis of the measuring results is carried out and the standard deviation of the contact angle are 0.86° for different droplet sizes and 0.93° for different curvature radii of the substrate. The analysis results demonstrate the stability and grid-independent of the present method to measure contact angle on curved wetting surfaces.

\subsection{\label{sec4.3}Deformations of sessile and pendent drops}
\begin{figure}[]
	\centering
		\includegraphics[scale=.75]{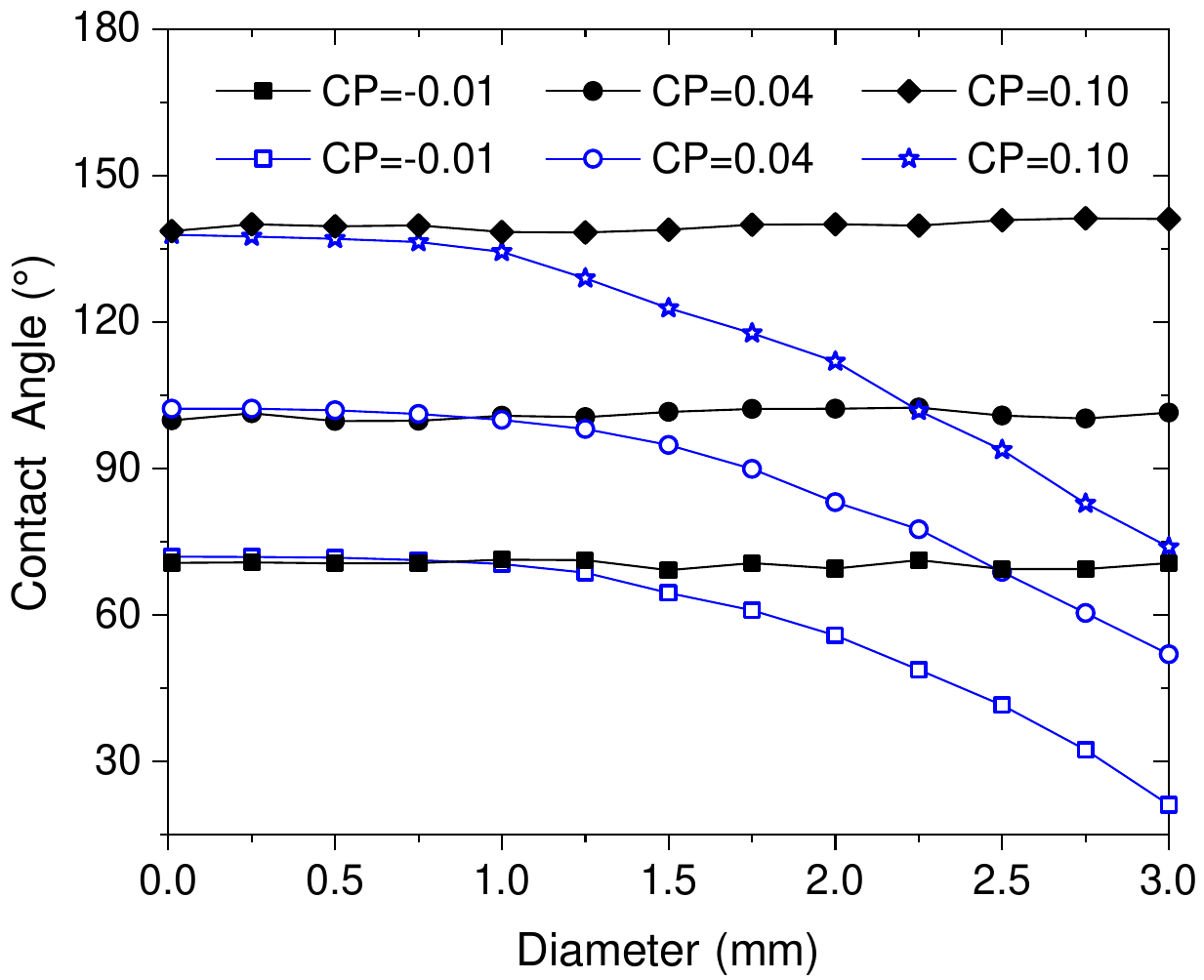}
	\caption{Contact angles of sessile droplets of different size on three solid surfaces. The droplet diameters vary from 0 mm to 3 mm, and the contact angles of the solid surfaces take 70$^{\circ}$, 100$^{\circ}$ and 140$^{\circ}$. The black solid symbols are the results of the present scheme and the blue hollow symbols are the results calculated by the spherical cap method.}
	\label{FIG:5}
\end{figure}

\begin{figure}[]
	\centering
		\includegraphics[]{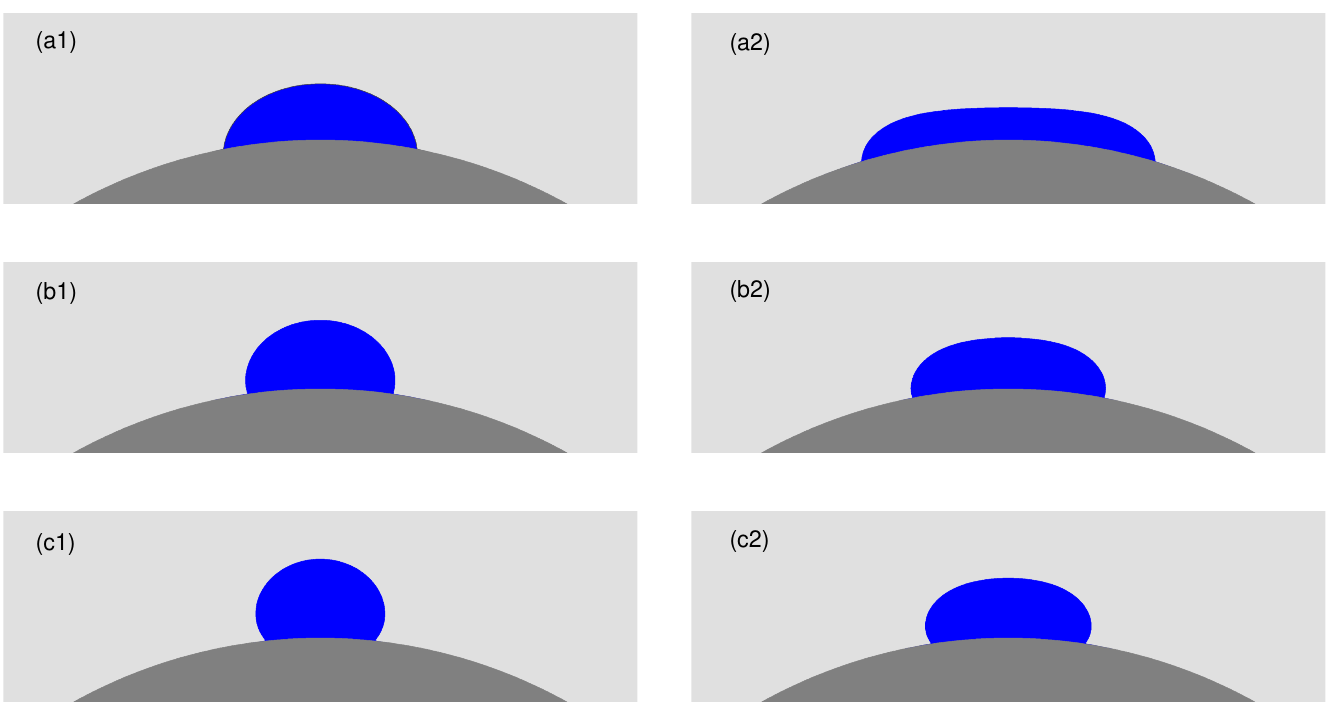}
	\caption{Deformation of sessile droplets of different sizes on three solid surfaces. The initial drop diameters are 2 mm for the left droplets and 3 mm for the right droplets. The contact angles of the solid surfaces are (a) 70$^{\circ}$, (b) 100$^{\circ}$ and (c) 140$^{\circ}$.}
	\label{FIG:6}
\end{figure}

Both theory and experiment have verified that gravity does not affect the equilibrium contact angle of a droplet on a uniformly smooth surface \cite{Lubarda2012} . If the effect of gravity is considered, the droplets on the solid surface will undergo deformation gradually deviating from the spherical cap shape. As the radius of the droplet increases, the effect of gravity becomes significant gradually, and the drop deformation is larger and larger. Because the deformation extends the drop footprint radius and lower the drop height, the contact angle computed by the spherical method, which is based on the height and footprint radius, decreases inevitably under gravity. A series of droplets are simulated and their diameters vary from 0 mm to 3 mm. These droplets were located on curved wetting substrates with the contact angles 70$^{\circ}$, 100$^{\circ}$ and 140$^{\circ}$, and the corresponding chemical potentials took -0.01, 0.04, and 0.1, respectively. The macroscopic diameter 0 is equivalent to the case of zero gravity. Fig. \ref{FIG:5} presents that both of the present scheme and the spherical cap method obtain almost the same contact angles when the droplet diameter is less than 1 mm. This confirms the theoretical prediction by Picknett and Bexon that a droplet resting on a smooth homogeneous surface takes the shape of a spherical cap and the gravity effect is negligible provided that its mass is less than about 1 mg \cite{Picknett1977}. When the macroscopic diameter of the droplet exceeds 1 mm, the contact angle calculated using the spherical cap method decreases significantly due to a decrease in height and an increase in width, and gradually deviates from the gravity-free value. However, the contact angles measured by the present method is remain the same all the time. This confirms that the present scheme obtains the microscopic contact angle, which is independent of gravity \cite{Lubarda2012}. The deformations of droplets under gravity are shown in Fig. \ref{FIG:6}. The macroscopic diameters of these two sets of droplets are 2 and 3 mm. Since the surface tensor of the water/vapor system is constant, the larger droplet suffers a larger gravity force and displays more apparent deformation. It can be clearly seen in Fig. \ref{FIG:6} that the droplets with the initial droplet diameter of 3 mm are much flatter and shorter than those of 2 mm. The simulation results are consistent with the those in the literature by Xie \emph{et al.} \cite{Xie2016}.

A pendent droplet adsorbed on the undersurface of a homogeneous curved substrate is stretched and its footprint radius is contacted by the gravity effect, thus the contact angle calculated by the spherical cap method inevitably increases with the growth of the droplet. A series of pendent droplets with the diameters varying from 0 mm to 2.25 mm are simulated, and the larger droplet will quickly detach from the substrate and fall off. The droplet with the macroscopic diameter 0 is equivalent to the case of zero gravity. The curved wetting surfaces have the contact angle 70$^{\circ}$, 100$^{\circ}$ and 140$^{\circ}$, and the corresponding chemical potentials are -0.01, 0.04, and 0.1, respectively. Fig. \ref{eq(7)} shows that on the same homogeneous surface, the contact angles of the droplets with different macroscopic diameters remain the same all the time, whereas those calculated by the spherical cap method gradually increase as the growth of the drops. Similar to the sessile drops, the independence between the microscopic contact angle and the gravity effect is verified again.
\begin{figure}[]
	\centering
		\includegraphics[scale=.75]{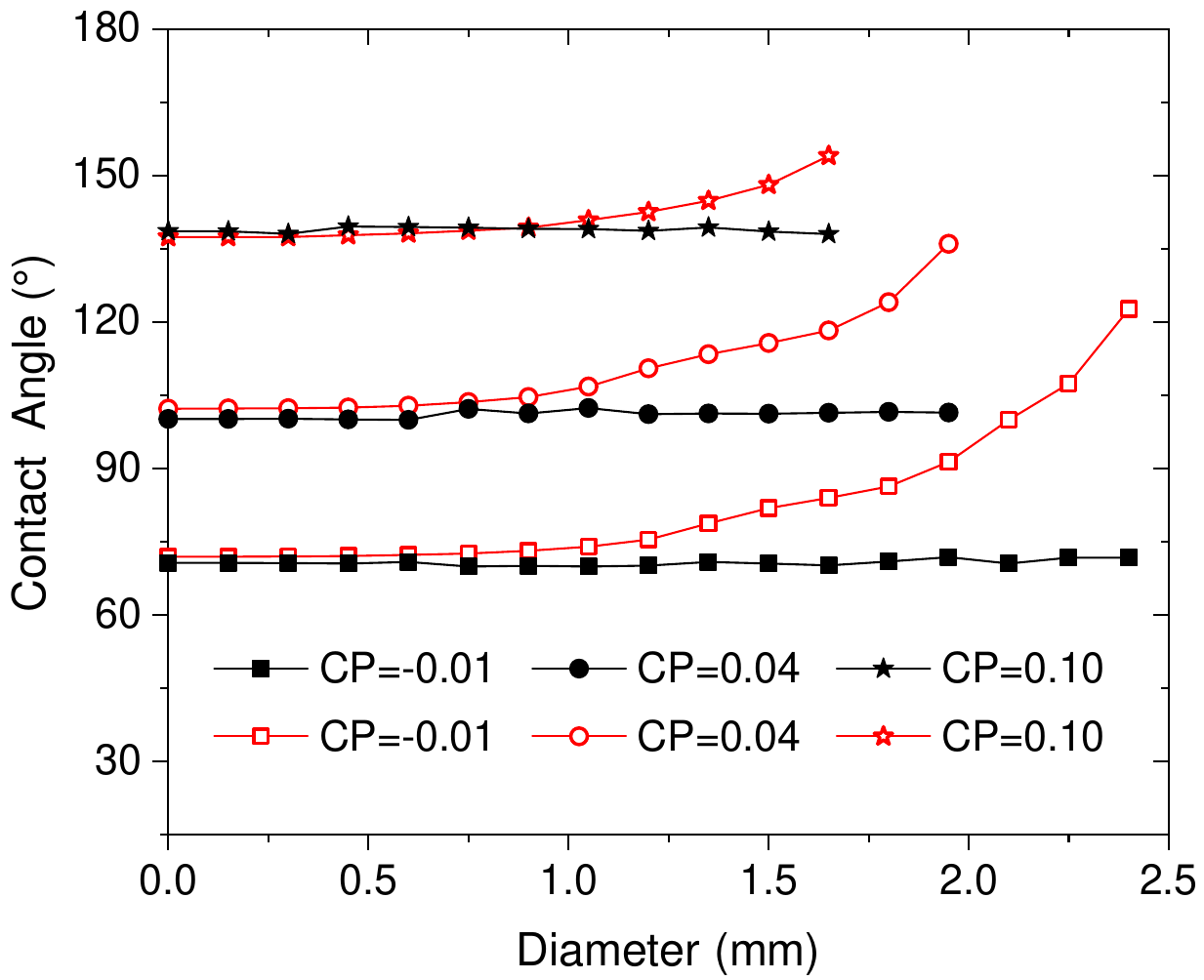}
	\caption{Contact angles of pendent droplets of different sizes on three solid surfaces. The droplet diameters vary from 0 mm to 2.5 mm, and the contact angles of the solid surfaces take 70$^{\circ}$, 100$^{\circ}$, 140$^{\circ}$. The black solid symbols are the results of the present scheme and the red hollow symbols are results calculated by the spherical cap method.}
	\label{FIG:7}
\end{figure}
\begin{figure}[]
	\centering
		\includegraphics[]{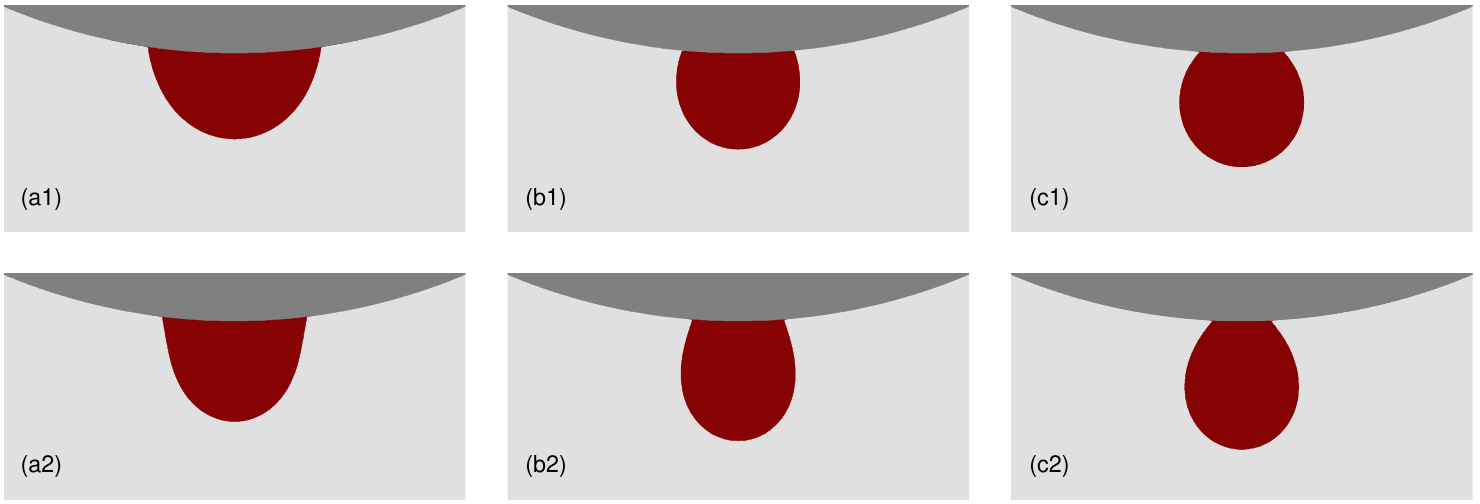}
	\caption{Deformation of pendent droplets of different diameters on three solid surfaces. The contact angles of the curved wetting substrate are (a) 70$^{\circ}$, (b) 100$^{\circ}$, (c) 140$^{\circ}$. The initial droplet diameters are (a1) 1.75, (a2) 2, (b1) 1.5, (b2) 1.75, (c1) 1.25 and (c2) 1.5 mm.}
	\label{FIG:8}
\end{figure}

The obvious difference between a pendent droplet and a sessile droplet is that the pendent droplet will fall off, when its size is large enough so that the gravity force is greater than the adhesion force. Therefore, as shown in Fig. \ref{FIG:8}, droplets cannot be stretched unceasingly. When the substrate is more hydrophobic, the pendent droplet is stretched and the footprint radius is narrower. The droplet falls off when the diameter exceeds 2.4, 1.95, 1.65 mm for the substrates with the contact angle 70$^{\circ}$, 100$^{\circ}$ and 140$^{\circ}$, respectively. 

\subsection{\label{sec4.4}Dynamic contact angle hysteresis}
\begin{figure}[]
	\centering
		\includegraphics[scale=.5]{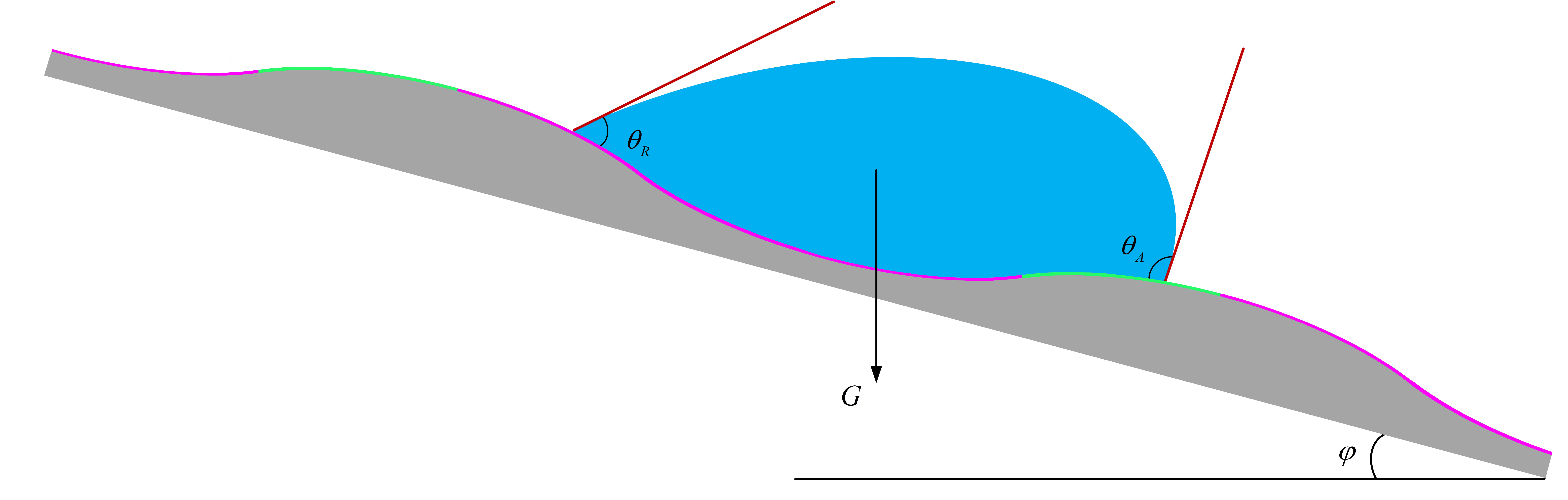}
	\caption{A schematic diagram to illustrate a droplet located on a chemically heterogeneous curved substrate with a slope angle $\varphi$. The segments in red and green represent hydrophobic and hydrophilic surfaces, respectively. With the effect of gravity $G$, the droplet displays deforming and moving, and the contact angle divide into an advancing angle ${\theta _A}$ and a receding angle ${\theta _R}$.}
	\label{FIG:9}
\end{figure}

In general, the roughness and chemical heterogeneity of a solid surface can lead to contact angle hysteresis \cite{Snoeijer2013}. In this section, we investigate the dynamic contact angle hysteresis on chemically homogeneous and heterogeneous curved surfaces, which is too difficult to capture in experiments and theoretical calculations. As shown in Fig. \ref{FIG:9}, the contact angles of the left and right sides of the droplet are no longer the same under the influence of gravity and the slope angle $\varphi $, which are called the advancing and receding contact angles ( ${\theta _A}$ and ${\theta _R}$ ). The spherical cap method is no longer applicable due to the deformation of the droplet during the motion. The initial drop radius is 100 lattice units, and its macroscopic diameter is 0.4 cm. As the slope angle of the curved plate is gradually increased, the advancing angle grows and the receding angle reduces. Once the droplet destabilizes, the advancing or receding angles will leave the initial position. 

\begin{figure}[]
	\centering
		\includegraphics[scale=.75]{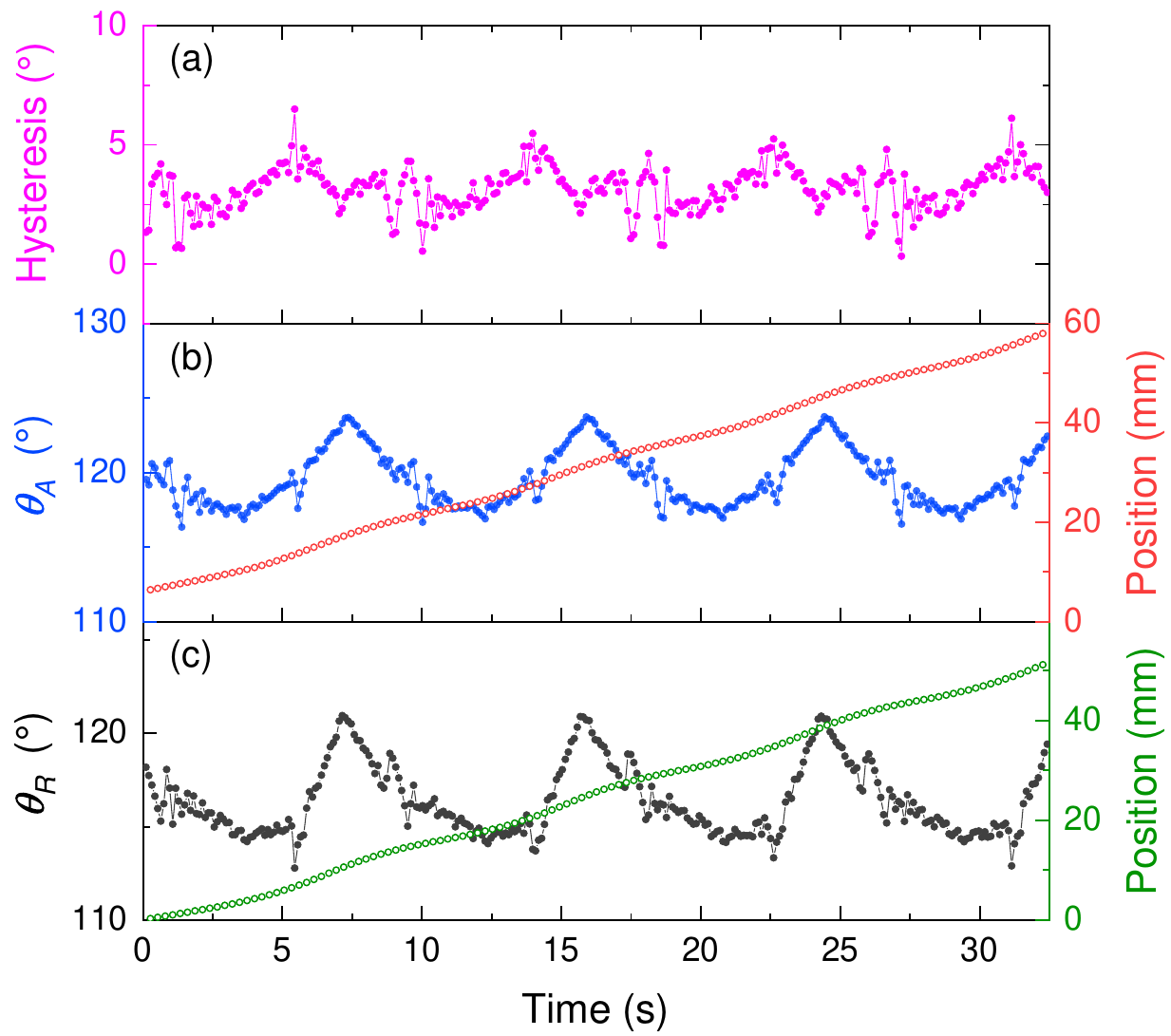}
	\caption{Movement of a droplet on a chemically homogeneous curved surface. (a) The dynamic contact angle hysteresis. (b) The advancing contact angle and its position.  (c) The receding contact angle and its position.}
	\label{FIG:10}
\end{figure}

Numerical simulations are firstly carried out on a homogeneous curved substrate with the contact angle 120° and the slope angle 20°. As shown in Fig. \ref{FIG:10}, the drop continuously moves on the substrate under the gravitational force. With the movements of the advancing and receding contact lines, the two contact angles periodically wave due to the curved geometry of the substrate, and the range is about 10°. The dynamic contact angle hysteresis indicates the difference between the advancing and receding contact angles in real time. The subfigure \ref{FIG:10}(a) plots the regular fluctuations. It is clear that the dynamic contact angle hysteresis caused by the homogeneous curved substrate is gentle and is basically limited in the range of 5°.

\begin{figure}[]
	\centering
		\includegraphics[scale=0.75]{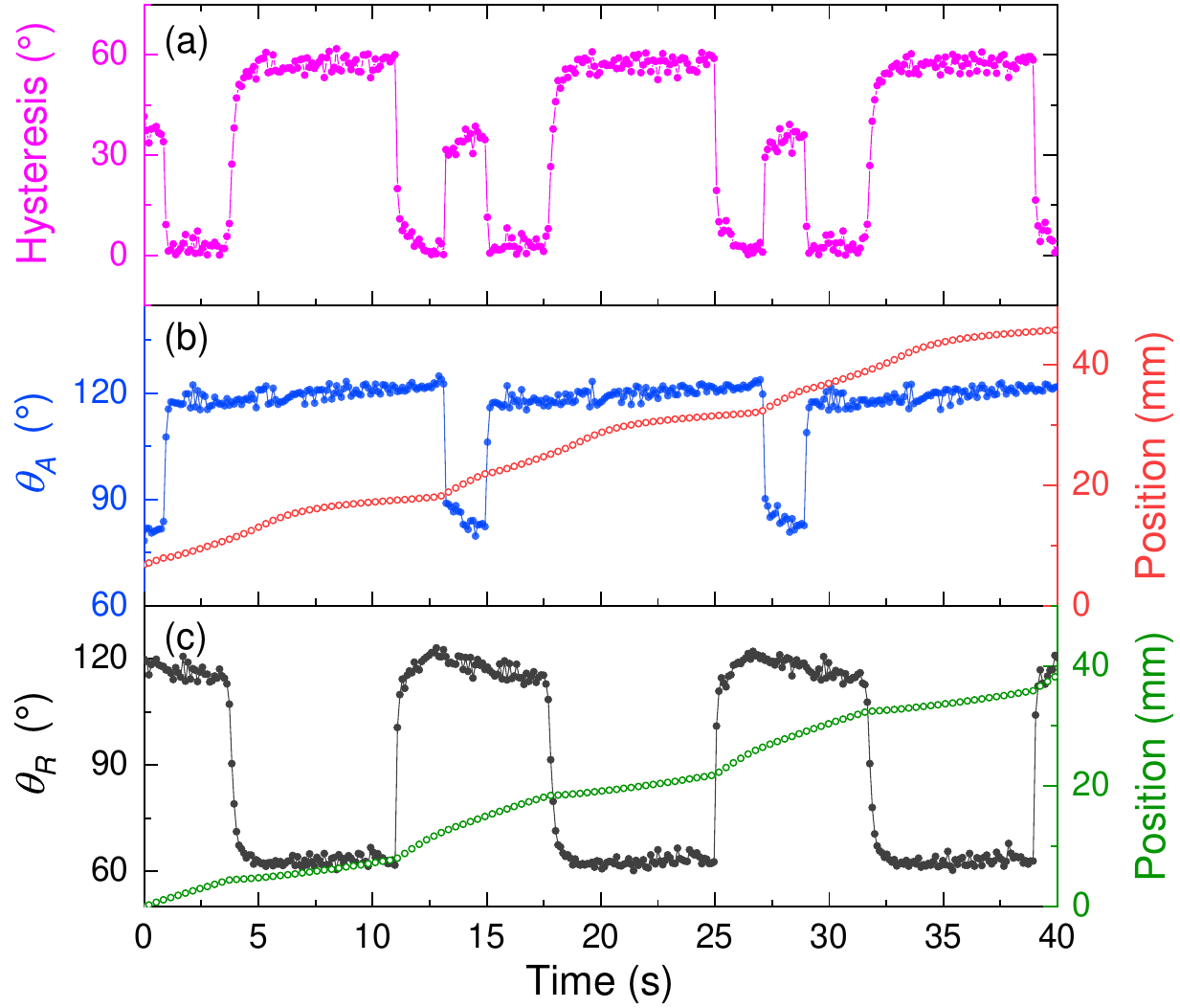}
	\caption{Movement of a droplet on a chemically heterogeneous curved surface. (a) The dynamic contact angle hysteresis. (b) The advancing contact angle and its position.  (c) The receding contact angle and its position.}
	\label{FIG:11}
\end{figure}

Then, the simulations apply a chemically heterogenous surface with the slope angle 20°, whose hydrophilic and hydrophobic regions have the contact angles 65° and 120° respectively. When the contact line crosses the border of the two regions, the clear stick-slip motions can be observed at the advancing and receding angles and produce the significant contact angle hysteresis. Fig. \ref{FIG:11} illustrates the continuous movement of the drop on the chemically heterogenous curved surface under the gravitational force. The changes of the advancing and receding angle are periodic and dramatical. Because they do not move synchronously, the dynamic contact angle hysteresis caused by the heterogenous substrate is very large and approach even 60°, as shown in the subfigure \ref{FIG:11}(a).

\begin{figure}[]
	\centering
		\includegraphics[scale=0.75]{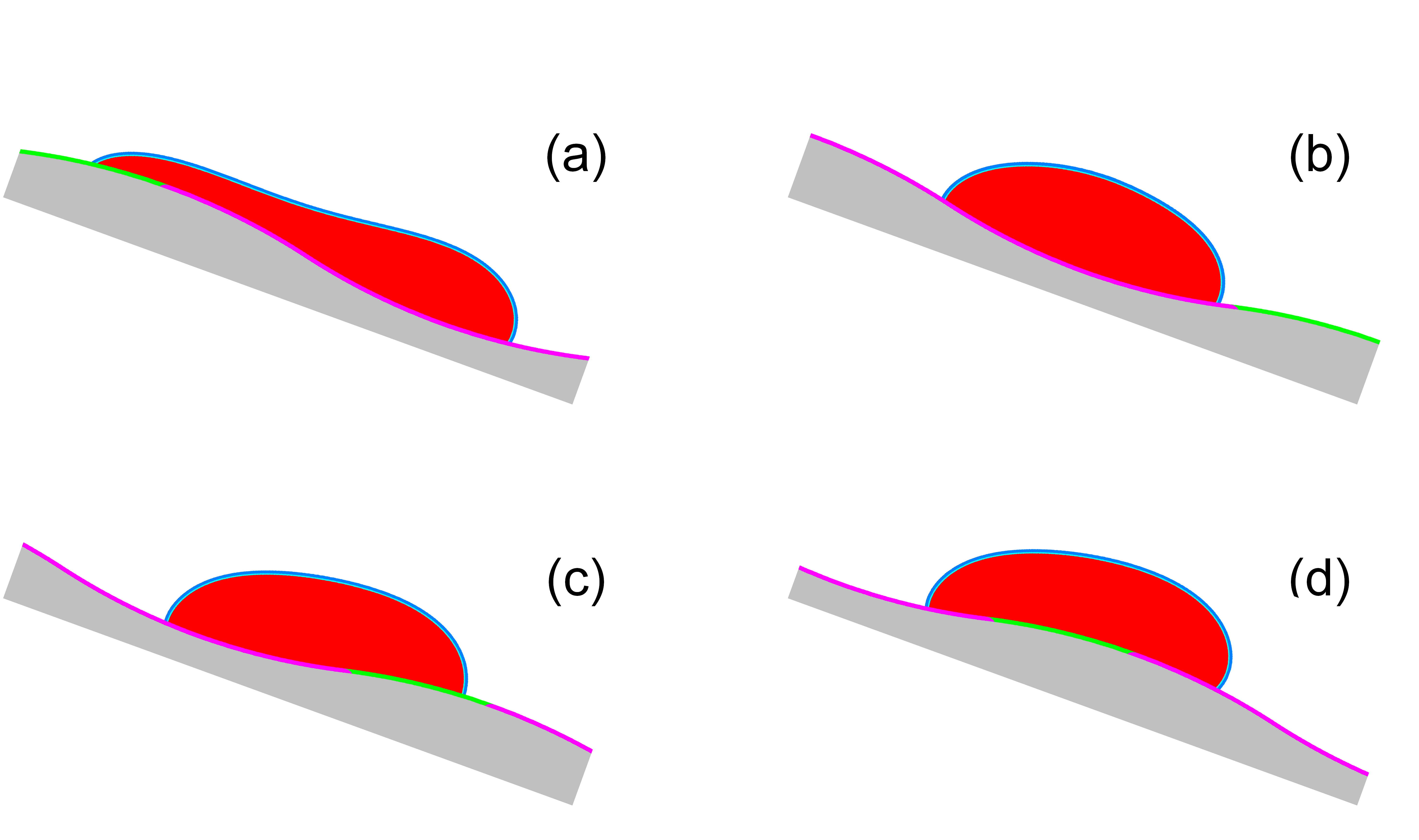}
	\caption{Snapshots of a drop moving on the chemically patterned curved surface at (a) t = 8 s, (b) t = 12 s, (c) t = 15 s, and (d) t = 17 s. The segments in red and green represent the hydrophobic and hydrophilic surfaces, respectively.}
	\label{FIG:12}
\end{figure}

Furthermore, the drop snapshots at four positions are drawn in Fig. \ref{FIG:12}. The drop in the subfigure (a) is elongated by the gravitational force, because the receding angle is hauled by the hydrophilic region. The subfigure (b) draws the drop is contracted, because it is on the hydrophobic region wholly. The subfigure (c) captures the advancing angle is fast spreading on the hydrophilic region. The hydrophilic region in the subfigure (d) is beneath the drop and has no influence on the contact angle; thus, as same as the subfigure (b), both the advancing and receding angle are about 120°. With the accurate contact angle measurement, one can readily conduct the in-situ mechanical analysis for every time steps. 

\subsection{\label{sec4.5}Mechanical analysis at moving contact line}
Fig. \ref{FIG:11} and \ref{FIG:12} present the fluctuations of the contact angle and the deformations of the drop shape. All these changes are related the force balance at the three-phase contact region. The present contact angle measurement enables the locally mechanical analysis in real time. The fluctuating contact angle results in an unbalanced Young’s force, which (per unit length) can be expressed as \cite{Sui2014} 
\begin{equation}
   F = \gamma \left( {\cos \theta  - \cos {\theta _{{\rm{eq}}}}} \right)\label{eq(17)}
\end{equation}                                                  
where  $\gamma $ represents the liquid-gas surface tension, and $\theta $ is the dynamic contact angle and  ${\theta _{eq}}$ is the equilibrium contact angle in relation to the surface property. When the liquid/gas transition region is on a homogeneous region, ${\theta _{eq}}$ is equal to the contact angle of the region. If it locates on a heterogeneous region that is composed of two components, the equilibrium contact angle is evaluated by the modified Cassie-Baxter equation \cite{Choi2009} :
\begin{equation}
   \cos {\theta _{{\rm{eq}}}} = r{\varphi _{\rm{d}}}\cos {\theta _{{\rm{s}}1}} + \left( {1 - {\varphi _{\rm{d}}}} \right)\cos {\theta _{{\rm{s}}2}}\label{eq(18)}
\end{equation} 
where  ${\theta _{s1}}$ and ${\theta _{s2}}$  are the intrinsic equilibrium contact angles for the two components, $r$ represents the roughness of the wetting surface ($r = 1$ for a smooth surface). ${\varphi _d}$  and  $1 - {\varphi _d}$ indicate the area ratios of the liquid/solid and liquid/gas interfaces. In the context of two-dimensional diffuse interface model, ${\varphi _d}$  and $1 - {\varphi _d}$  refer to the ratio of the length of the isodensity line on the surfaces with contact angles of ${\theta _{s1}}$  and ${\theta _{s2}}$ , respectively, to the total length of the isodensity lines. When the contact line is located at the border of hydrophilic and hydrophobic surfaces, we use Eq. \eqref{eq(18)} to calculate the corresponding equilibrium contact angle.
\begin{figure}[]
	\centering
		\includegraphics[scale=0.75]{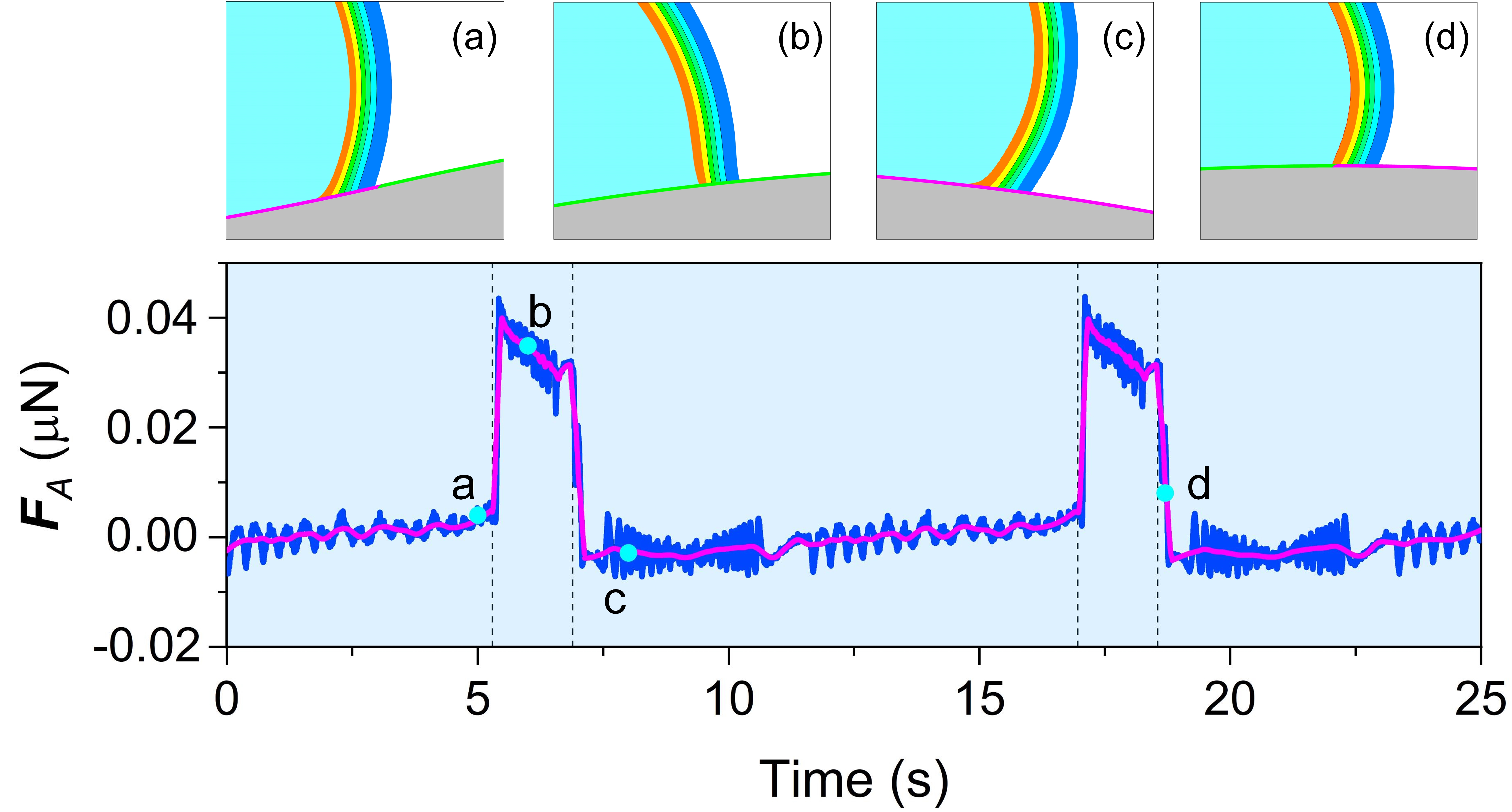}
	\caption{The unbalanced Young’s force at the advancing angle during the drop movement on the chemically patterned curved surface. The blue line is the unbalanced Young’s force, and the purple line is the smoothed results. The insets draw the snapshots of the advancing angle at (a) t = 5 s, (b) t = 6 s, (c) t = 8 s, and (d) t = 18 s, which are also marked by the cyan dot on the force evolution. The dash lines indicate the moments that the advancing contact line crosses the border of hydrophilic and hydrophobic regions.}
	\label{FIG:13}
\end{figure}

Fig. \ref{FIG:13} presents the unbalanced Young’s force at the advancing angle during the stick-slip of a drop on a chemically heterogeneous surface, and the black dashed lines indicate the moments that the contact line crosses the border. The advancing angle in the hydrophobic region during the stick-slip movement fluctuates around the equilibrium angle, so the unbalanced Young's force also fluctuates around zero, as shown in Fig. \ref{FIG:13}. The force exhibits a large jump when the advancing contact line crosses the border of hydrophilic and hydrophobic regions. In the hydrophilic region, the advancing angle is greater than the equilibrium angle because the hydrophobic surface resists the liquid spreading; thus, the Young’s force on this region is much greater than zero.
\begin{figure}[]
	\centering
		\includegraphics[scale=0.75]{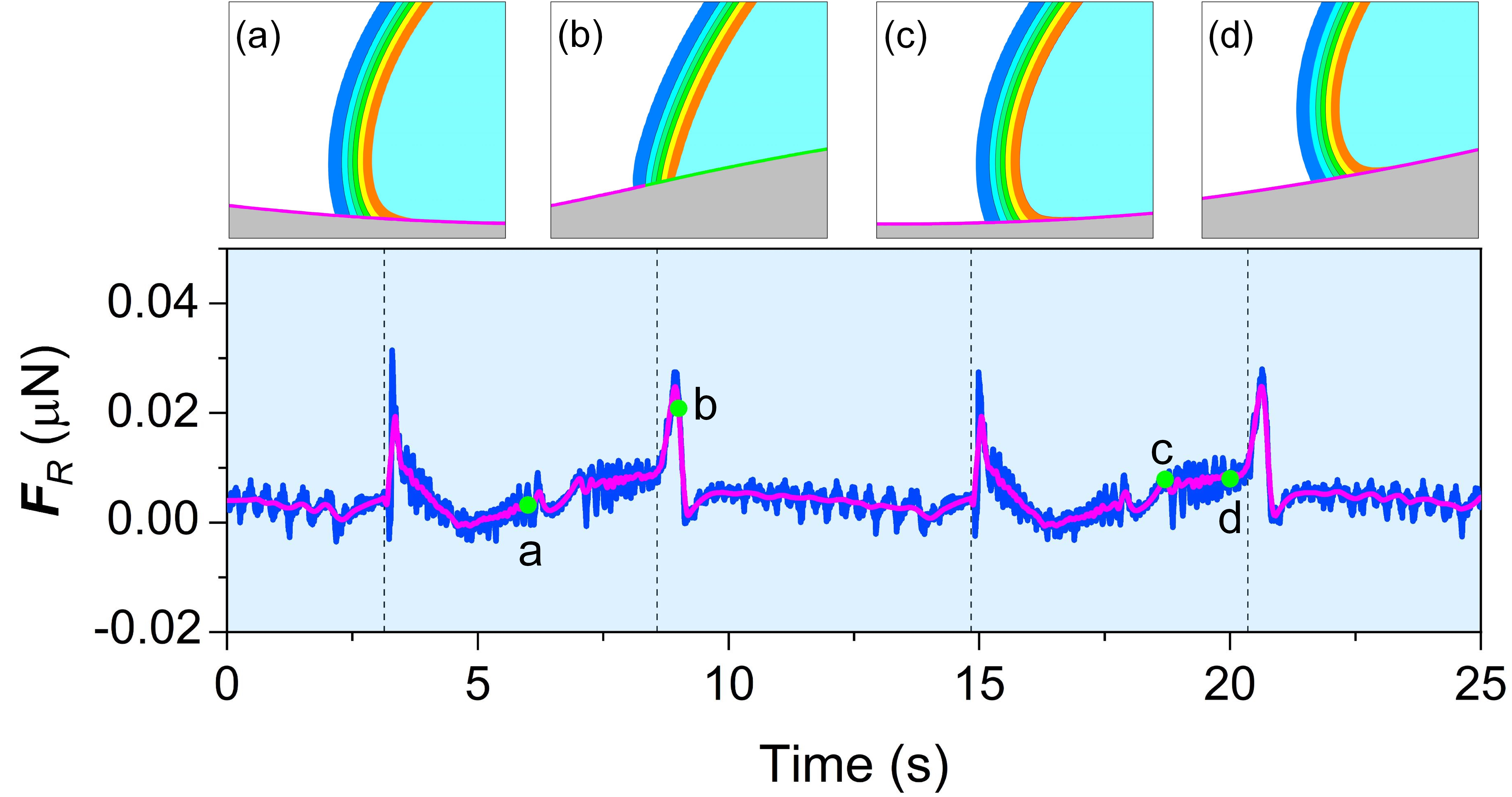}
	\caption{The unbalanced Young’s force at the receding angle during the drop movement on the chemically patterned curved surface. The blue line is the unbalanced Young’s force, and the purple line is the smoothed results. The insets draw the snapshots of the receding angle at (a) t = 6 s, (b) t = 9 s, (c) t = 18 s, and (d) t = 20 s, which are also marked by the green dots on the force evolution. The dash lines indicate the moments that the receding contact line crosses the border of hydrophilic and hydrophobic regions}
	\label{FIG:14}
\end{figure}

Fig. \ref{FIG:14} presents the unbalanced Young’s force of the receding angle at the stick-slip motion of a drop on a chemically heterogeneous surface. The force in the hydrophilic region oscillates and is often greater than zero. When the receding contact line crosses the border of hydrophilic and hydrophobic regions, the force also exhibits a large jump. It should be note that since the right is the positive direction,  $\theta_A  > {\theta _{eq}}$ leads to a positive unbalanced Young’s force, whereas  $\theta_R  > {\theta _{eq}}$ leads to a negative one. The simulation results manifest that the present scheme can accurately evaluate the dynamic contact angle and conduct the in-situ mechanical analysis at moving contact line. We expect to gain further insight into capillary phenomenon and dynamic hysteresis through microscopic contact angle and real-time mechanical analysis.
\section{\label{sec5}Conclusions}
Wetting and capillarity are ubiquitous in nature. As the most important physical quantity in this field, contact angle illustrates the competitive interactions between liquid, gas and solid surface. It is significant to obtain the accurate contact angle in scientific researches and industrial applications. This paper presents a geometry-based scheme to measure the real-time contact angle on curved wetting substrates. The accuracy and gird independence of the scheme are carefully verified. The theoretical prediction that the microscopic contact angle does not depend on gravity is confirm, even though drops are deformed under gravity. With the accurate measurement of microscopic contact angle, the dynamic contact angle hysteresis can be captured readily, and the mechanical analyses at moving contact line are implemented in situ. These simulations manifest that the present scheme can be a powerful tool to investigate the issues involving surface wetting, capillary phenomena and moving contact lines. Since it is based on the geometry of contact angle, the present scheme can be used in other multiphase lattice Boltzmann models, such as pseudopotential models \cite{Chen2014,Li2016,Shan1993}, field phase models \cite{Zheng2006,Liang2018}, etc.

\begin{acknowledgments}
This work was supported by the National Natural Science Foundation of China (Grant Nos. 11862003, 81860635, and 12062005), the Key Project of Guangxi Natural Science Foundation (Grant No. 2017GXNSFDA198038), Guangxi “Bagui Scholar” Teams for Innovation and Research Project, and Guangxi Collaborative Innovation Center of Multisource Information Integration and Intelligent Processing. 
\end{acknowledgments}

%

\end{document}